\documentstyle[12pt]{article}
\def\1{{\chi}}

\begin{document}
\title {{Generalized quantum operations and almost sharp quantum effects}\thanks{This project is supported by Natural Science
Found of China (10771191 and 10471124).}}
\author {Shen Jun$^{1,2}$, Wu Junde$^{1}$\date{}\thanks{E-mail: wjd@zju.edu.cn}}
\maketitle
$^1${\small\it Department of Mathematics, Zhejiang
University, Hangzhou 310027, P. R. China}

$^2${\small\it Department of Mathematics, Anhui Normal University,
Wuhu 241003, P. R. China}

\begin{abstract} {\noindent In this paper, we study generalized quantum operations and almost sharp quantum effects, our results generalize and improve some important conclusions in [2] and [3].
}
\end{abstract}


\vskip 0.2 in

\noindent {\bf This paper is to commemorate my outstanding student
Shen Jun, who passed away accidently on July 1, 2009. Shen Jun made
great contributions in sequential effect algebra theory. He solved
four open problems which were presented by Professor Gudder in
International Journal of Theoretical Physics, 44 (2005), 2199-2205.}

\vskip 0.2 in

{\bf 1. Introduction}

\vskip 0.2 in

\noindent Let $H$ be a Hilbert space, $B(H)$ be the set of bounded
linear operators on $H$, $P(H)$ be the set of projection operators
on $H$, $T(H)$ be the set of trace class operators on $H$, and
$\Gamma=\{A_\alpha,A_\alpha^*\}_{\alpha\in \Lambda}$ be a set of
operators, where $A_\alpha\in B(H)$ satisfy $\sum\limits_\alpha
A_\alpha A_\alpha^*\leq I$. A map $\Phi_{\Gamma}:B(H)\longrightarrow
B(H); B\longmapsto \sum\limits_\alpha A_\alpha B A_\alpha^*$ is
called a {\it generalized quantum operation}. Each element of
$\Gamma=\{A_\alpha,A_\alpha^*\}_{\alpha\in \Lambda}$ is said to be a
operation element  of $\Phi_{\Gamma}$. If $B\geq 0$, then it is
obvious that $\sum\limits_\alpha A_\alpha B A_\alpha^*$ converges in
the strong operator topology, so $\sum\limits_\alpha A_\alpha B
A_\alpha^*$ converges in the strong operator topology for any $B\in
B(H)$. If $\Phi_{\Gamma}(I)=\sum\limits A_\alpha A^*_\alpha=I$, then
$\Phi_{\Gamma}$ is said to be
 {\it unital}, if $\sum\limits_\alpha A_\alpha^* A_\alpha = I$, then $\Phi_{\Gamma}$
 is said to be {\it trace preserving}, if $\sum\limits_\alpha A_\alpha^* A_\alpha \leq I$, then $\Phi_{\Gamma}$
 is said to be {\it trace nonincreasing}, if $A_\alpha^*=A_\alpha$ for every $\alpha$, then $\Phi_{\Gamma}$
 is said to be {\it self-adjoint}.

\vskip 0.1 in

The set of fixed points of $\Phi_{\Gamma}$ is
$B(H)^{\Phi_{\Gamma}}=\{B\in B(H)\mid \Phi_{\Gamma}(B)=B\}$.
Obviously $B(H)^{\Phi_{\Gamma}}$ is closed under the involution $*$.
The commutant ${\Gamma}'=\{B\in B(H)\mid BA_\alpha=A_\alpha B,
BA_\alpha^*=A_\alpha^* B, \alpha\in \Lambda\}$ of ${\Gamma}$ is a
von Neumann algebra.

\vskip 0.1 in

Quantum operations frequently occur in quantum measurement theory,
quantum probability, quantum computation, and quantum information
theory ([1]). If an operator $A$ is invariant under the quantum
operation $\Phi_{\Gamma}$, in physics, it implies that $A$ is not
disturbed by the action of $\Phi_{\Gamma}$. So, the following
problem is interesting and important: if $A$ is a
$\Phi_{\Gamma}$-fixed point, is $A$ commutative with each operation
element of $\Phi_{\Gamma}$? In general, the answer is not and some
sufficient conditions under which the answer is yes were given
([2]).

\vskip 0.1 in

On the other hand, quantum effects are represented by operators on a
Hilbert space $H$ satisfying that $0\leq A\leq I$, and sharp quantum
effects are represented by projections. An quantum effect $A$ is
said to be almost sharp if $A=PQP$ for projections $P$ and $Q$
([3]). In [3], some characterizations of almost sharp quantum
effects were obtained.

\vskip 0.1 in

In this paper, we generalize some theorems in [2] from quantum
operations to generalized quantum operations, from unital to not
necessarily unital, and from trace preserving to trace
nonincreasing, we also generalize some results in [3] and give some
more characterizations for almost sharp quantum effects.

\vskip 0.2 in

{\bf 2. Generalized quantum operations}

\vskip 0.2 in

\vskip 0.1 in

{\bf Lemma 2.1.} If $\Phi_{\Gamma}$ is a generalized quantum
operation, $B,BB^*\in B(H)^{\Phi_{\Gamma}}$, then
$BA_\alpha=A_\alpha B$ for every $\alpha$.

{\bf Proof.} Since $B\in B(H)^{\Phi_{\Gamma}}$, we have $B^*\in
B(H)^{\Phi_{\Gamma}}$. Let we denote $
[B,A_\alpha]=BA_\alpha-A_\alpha B$. Note that $0\leq
[B,A_\alpha][B,A_\alpha]^*=(BA_\alpha-A_\alpha
B)(A_\alpha^*B^*-B^*A_\alpha^*)=BA_\alpha A_\alpha^*B^*+A_\alpha
BB^*A_\alpha^*-A_\alpha BA_\alpha^*B^*-BA_\alpha B^*A_\alpha^*$.

Thus $0\leq \sum\limits_\alpha
[B,A_\alpha][B,A_\alpha]^*=B(\sum\limits_\alpha A_\alpha
A_\alpha^*)B^*+\Phi_{\Gamma}(BB^*)-\Phi_{\Gamma}(B)B^*-B\Phi_{\Gamma}(B^*)=B(\sum\limits_\alpha
A_\alpha A_\alpha^*)B^*-BB^*\leq 0$.

So we conclude that $[B,A_\alpha]=0$ for every $\alpha$. That is,
$BA_\alpha=A_\alpha B$ for every $\alpha$.

\vskip 0.1 in

{\bf Theorem 2.1.} If $\Phi_{\Gamma}$ is a generalized quantum
operation, $B,B^*B,BB^*\in B(H)^{\Phi_{\Gamma}}$, then $B\in
{\Gamma}'$.

{\bf Proof.} By Lemma 2.1, $BA_\alpha=A_\alpha B$ for every
$\alpha$. Since $B\in B(H)^{\Phi_{\Gamma}}$, we have $B^*\in
B(H)^{\Phi_{\Gamma}}$. Thus by Lemma 2.1 again,
$B^*A_\alpha=A_\alpha B^*$ for every $\alpha$. Taking adjoint, we
have $BA_\alpha^*=A_\alpha^* B$ for every $\alpha$. So we conclude
that $B\in {\Gamma}'$.

\vskip 0.1 in

{\bf Theorem 2.2.} If $\Phi_{\Gamma}$ is a self-adjoint generalized
quantum operation, $B,BB^*\in B(H)^{\Phi_{\Gamma}}$, then $B\in
{\Gamma}'$.

{\bf Proof.} By Lemma 2.1, $BA_\alpha=A_\alpha B$ for every
$\alpha$. Since $A_\alpha^*=A_\alpha$ for every $\alpha$, we
conclude that $B\in {\Gamma}'$.

\vskip 0.1 in

We denote the set of selfadjoint elements in $B(H)^{\Phi_{\Gamma}}$
by $Re(B(H)^{\Phi_{\Gamma}})$.

{\bf Theorem 2.3.} If $\Phi_{\Gamma}$ is a generalized quantum
operation, then the following conditions are all equivalent:

(1) $B(H)^{\Phi_{\Gamma}}\subseteq {\Gamma}'$;

(2) If $B\in B(H)^{\Phi_{\Gamma}}$, then $B^*B\in
B(H)^{\Phi_{\Gamma}}$;

(3) If $B\in Re(B(H)^{\Phi_{\Gamma}})$, then $B^2\in
B(H)^{\Phi_{\Gamma}}$.

{\bf Proof.} (1)$\Rightarrow$(2): If $B\in B(H)^{\Phi_{\Gamma}}$,
then $B\in {\Gamma}'$. Thus $B^*\in {\Gamma}'$. So
$\Phi_{\Gamma}(B^*B)=\sum\limits_\alpha A_\alpha B^*B
A_\alpha^*=B^*\sum\limits_\alpha A_\alpha B
A_\alpha^*=B^*\Phi_{\Gamma}(B)=B^*B$. Thus $B^*B\in
B(H)^{\Phi_{\Gamma}}$.

(2)$\Rightarrow$(3) is obvious.

(3)$\Rightarrow$(1): By Theorem 2.1, If $B\in
Re(B(H)^{\Phi_{\Gamma}})$, then $B\in {\Gamma}'$. That is,
$Re(B(H)^{\Phi_{\Gamma}})\subseteq {\Gamma}'$. Since
$B(H)^{\Phi_{\Gamma}}$ is closed under the involution $*$, we
conclude that $B(H)^{\Phi_{\Gamma}}\subseteq {\Gamma}'$.

\vskip 0.1 in

{\bf Lemma 2.2.} If $\{C_\beta \}_{\beta}\subset B(H)$, $\{C_\beta
\}_\beta$ is a nondecreasing net of positive operators converging to
some $C_0\in B(H)$ in the strong operator topology, then
$tr(C_\beta)\longrightarrow tr(C_0)$, here the trace function
$tr(\cdot)$ can take value $+\infty$.

{\bf Proof.} Since $0\leq C_\beta\leq C_0$, we have $tr(C_\beta)\leq
tr(C_0)$.

For any constant $\xi< tr(C_0)=\sum\limits_{\gamma\in F}\langle
C_0x_\gamma,x_\gamma\rangle$ ( here $\{x_\gamma \}_{\gamma\in F}$ is
an orthonormal  bases of $H$), there exists a finite subset
$F_0\subseteq F$ such that $\xi<\sum\limits_{\gamma\in F_0}\langle
C_0x_\gamma,x_\gamma\rangle$. Since $\sum\limits_{\gamma\in
F_0}\langle C_\beta x_\gamma,x_\gamma\rangle\longrightarrow
\sum\limits_{\gamma\in F_0}\langle C_0x_\gamma,x_\gamma\rangle$, we
have $tr(C_\beta)\geq \sum\limits_{\gamma\in F_0}\langle C_\beta
x_\gamma,x_\gamma\rangle> \xi$ for all sufficiently large $\beta$.
Thus $tr(C_\beta)\longrightarrow tr(C_0)$.

\vskip 0.1 in

{\bf Theorem 2.4.} Let $\Phi_{\Gamma}$ be a trace nonincreasing
generalized quantum operation, $B\in T(H)_+$, then
$\Phi_{\Gamma}(B)\in T(H)_+$ and $tr(\Phi_{\Gamma}(B))\leq tr(B)$.

{\bf Proof.} Let $F$ be a finite subset of $\Lambda$, then
$tr(\sum\limits_{\alpha\in F} A_\alpha B
A_\alpha^*)=tr(\sum\limits_{\alpha\in F} A_\alpha^*A_\alpha B )\leq
\parallel \sum\limits_{\alpha\in F} A_\alpha^*A_\alpha\parallel
tr(B)\leq tr(B)$. Ordering all such $F$ by including,
$\{\sum\limits_{\alpha\in F} A_\alpha B A_\alpha^*\}_F$ is a
nondecreasing net of positive operators converging to
$\Phi_{\Gamma}(B)$ in the strong operator topology. So by Lemma 2.2
we have $tr(\sum\limits_{\alpha\in F} A_\alpha B
A_\alpha^*)\longrightarrow tr(\Phi_{\Gamma}(B))$. Thus
$tr(\Phi_{\Gamma}(B))\leq tr(B)$.

\vskip 0.1 in

A generalized quantum operation $\Phi_{\Gamma}$ is {\it faithful} if
for any $B\in B(H)$, $\Phi_{\Gamma}(B^*B)=0$ implies $B=0$.

{\bf  Theorem 2.5.} Let $\Phi_{\Gamma}$ be a trace preserving
generalized quantum operation. We have

(1). $\Phi_{\Gamma}$ is faithful.

(2). If $B\in T(H)$, then $\Phi_{\Gamma}(B)\in T(H)$ and
$tr(\Phi_{\Gamma}(B))=tr(B)$.

{\bf Proof.} (1). Suppose $B\in B(H)$, $\Phi_{\Gamma}(B^*B)=0$. Then
$\sum\limits_\alpha A_\alpha B^*B A_\alpha^*=0$. So $B A_\alpha^*=0$
for every $\alpha$. Thus $B=B\sum\limits_\alpha A_\alpha^*
A_\alpha=0$.

(2). Firstly we suppose $B\in T(H)_+$. By Theorem 2.4 we have
$\Phi_{\Gamma}(B)\in T(H)_+$. Let $F$ be a finite subset of
$\Lambda$, ordering all such $F$ by including,
$\{\sum\limits_{\alpha\in F} A_\alpha B A_\alpha^*\}_F$ is a
nondecreasing net of positive operators converging to
$\Phi_{\Gamma}(B)$ in the strong operator topology. So by Lemma 2.2
we have $tr(\sum\limits_{\alpha\in F} A_\alpha B
A_\alpha^*)\longrightarrow tr(\Phi_{\Gamma}(B))$.

Since $\Phi_{\Gamma}$ is trace preserving, $\{\sum\limits_{\alpha\in
F} B^{\frac{1}{2}}A_\alpha^*A_\alpha B^{\frac{1}{2}} \}_F$ is a
nondecreasing net of positive operators converging to $B$ in the
strong operator topology. So by Lemma 2.2 we have
$tr(\sum\limits_{\alpha\in F} B^{\frac{1}{2}}A_\alpha^*A_\alpha
B^{\frac{1}{2}} )\longrightarrow tr(B)$. But
$tr(\sum\limits_{\alpha\in F} A_\alpha B
A_\alpha^*)=tr(\sum\limits_{\alpha\in F}
B^{\frac{1}{2}}A_\alpha^*A_\alpha B^{\frac{1}{2}} )$ for every $F$,
so we conclude that $tr(\Phi_{\Gamma}(B))=tr(B)$. By linearity, the
result for arbitrary $B\in T(H)$ now follows.

\vskip 0.1 in

The next Lemma 2.3 is from [4], it is presumed in [4] that all
linear maps on $C^*$-algebras preserve the identity, we modify the
proof slightly such that it suit for our need.

\vskip 0.1 in

{\bf Lemma 2.3. } If $\Re_1$, $\Re_2$ are $C^*$-algebras,
$\phi:\Re_1\longrightarrow\Re_2$ is a 2-positive linear map,
$\|\phi(I)\|\leq 1$, then $\phi(C^*C)\geq \phi(C)^*\phi(C)$ for
every $C\in\Re_1$.

{\bf Proof.} Let $T=\left(
                      \begin{array}{cc}
                        0 & C^* \\
                        C & 0 \\
                      \end{array}
                    \right)\in M_2(\Re_1)=\Re_1\otimes M_2
$, here $M_2$ denote the $C^*$-algebra of $2\times2$ complex
matrices. Then $T=T^*$.

Since $\phi\otimes 1_2:M_2(\Re_1)\longrightarrow M_2(\Re_2)$ is a
positive linear map and $\|\phi\otimes 1_2\|\leq 1$, by [5] Theorem
1 we have $(\phi\otimes 1_2)(T^2)\geq ((\phi\otimes 1_2)(T))^2$.

While $T^2=\left(
                      \begin{array}{cc}
                        C^*C & 0 \\
                        0 & CC^* \\
                      \end{array}
                    \right)
$, $(\phi\otimes 1_2)(T^2)=\left(
                      \begin{array}{cc}
                        \phi(C^*C) & 0 \\
                        0 & \phi(CC^*) \\
                      \end{array}
                    \right)
$,

$(\phi\otimes 1_2)(T)=\left(
                      \begin{array}{cc}
                        0 & \phi(C^*) \\
                        \phi(C) & 0 \\
                      \end{array}
                    \right)
$, $((\phi\otimes 1_2)(T))^2=\left(
                      \begin{array}{cc}
                        \phi(C)^*\phi(C) & 0 \\
                        0 & \phi(C)\phi(C)^* \\
                      \end{array}
                    \right)
$.

Thus $\phi(C^*C)\geq \phi(C)^*\phi(C)$.

\vskip 0.1 in

It is easy to see that a generalized quantum operation is completely
positive and satisfies the conditions in Lemma 2.3.

\vskip 0.1 in

An operator $W\in T(H)$ is {\it faithful} if for any $A\in B(H)_+$,
$tr(W^*AW)=0$ implies $A=0$.

\vskip 0.1 in

{\bf Theorem 2.6.} Let $\Phi_{\Gamma}$ be a trace nonincreasing
generalized quantum operation. We have

(1). $B(H)^{\Phi_{\Gamma}}\cap T(H)\subseteq{\Gamma}'\cap T(H)$;

(2). If $dim(H)<\infty$, then $B(H)^{\Phi_{\Gamma}}\subseteq
{\Gamma}'$;

(3). If there exists a faithful operator $W\in T(H)\cap {\Gamma}'$,
then $B(H)^{\Phi_{\Gamma}}\subseteq {\Gamma}'$.

{\bf Proof.} (1). Suppose $B\in B(H)^{\Phi_{\Gamma}}\cap T(H)$. Thus
$B^*B\in T(H)_+$. By Lemma 2.3 we have $\Phi_{\Gamma}(B^*B)\geq
 \Phi_{\Gamma}(B)^*\Phi_{\Gamma}(B)=B^*B$. By Theorem 2.4 we have $\Phi_{\Gamma}(B^*B)\in T(H)_+$ and
$tr(\Phi_{\Gamma}(B^*B))=tr(B^*B)$. That is,
$tr(\Phi_{\Gamma}(B^*B)-B^*B)=0$. So $\Phi_{\Gamma}(B^*B)=B^*B$. We
conclude that $B^*B\in B(H)^{\Phi_{\Gamma}}$. Since
$B(H)^{\Phi_{\Gamma}}$ is closed under the involution $*$, we also
have $B^*\in B(H)^{\Phi_{\Gamma}}\cap T(H)$. Similarly we have
$BB^*\in B(H)^{\Phi_{\Gamma}}$. By Theorem 2.1, We conclude that
$B\in {\Gamma}'$. That is, $B(H)^{\Phi_{\Gamma}}\cap T(H)\subseteq
{\Gamma}'\cap T(H)$.

(2) follows from (1) immediately.

(3). Suppose $B\in B(H)^{\Phi_{\Gamma}}$. By Lemma 2.3 we have
$\Phi_{\Gamma}(B^*B)\geq
 \Phi_{\Gamma}(B)^*\Phi_{\Gamma}(B)=B^*B$. Thus By Theorem 2.4 we
 have $$0\leq
tr(W^*(\Phi_{\Gamma}(B^*B)-B^*B)W)$$$$=tr(W^*\Phi_{\Gamma}(B^*B)W)-tr(W^*B^*BW)$$$$=tr(\Phi_{\Gamma}(W^*B^*BW))-tr(W^*B^*BW)\leq0.$$
So $tr(W^*(\Phi_{\Gamma}(B^*B)-B^*B)W)=0$. Since $W$ is faithful, we
conclude that $\Phi_{\Gamma}(B^*B)=B^*B$. That is, $B^*B\in
B(H)^{\Phi_{\Gamma}}$. Since $B(H)^{\Phi_{\Gamma}}$ is closed under
the involution $*$, we also have $B^*\in B(H)^{\Phi_{\Gamma}}$.
Similarly we have $BB^*\in B(H)^{\Phi_{\Gamma}}$. By Theorem 2.1, we
conclude that $B\in {\Gamma}'$. That is,
$B(H)^{\Phi_{\Gamma}}\subseteq {\Gamma}'$.

\vskip 0.1 in

The next theorem is a direct corollary of Theorem 2.6 (2), but we
give a simple elementary proof instead.

\vskip 0.1 in

{\bf Theorem 2.7.} Let $\Phi_{\Gamma}$ be a generalized quantum
operation, $\Gamma=\{A_\alpha,A_\alpha^*\}_{\alpha\in \Lambda}$ is
commutative and $dim(H)<\infty$, then $B(H)^{\Phi_{\Gamma}}\subseteq
{\Gamma}'$.

{\bf Proof.} By Theorem 2.5.5 in [6], $\{A_\alpha\}_{\alpha\in
\Lambda}$ can be diagonalized simultaneously. That is, there exists
a set of pairwise orthogonal nonzero projections $\{P_k\}_k$ such
that $\sum\limits_{k} P_k=I$, $A_\alpha=\sum\limits_{k}
\lambda_{k,\alpha}P_k$. We also can suppose that if $k_1\neq k_2$,
then there exists some $\alpha$ such that
$\lambda_{k_1,\alpha}\neq\lambda_{k_2,\alpha}$. In fact, if not, we
can combine $P_{k_1}$ and $P_{k_2}$ into one projection.

Since $\sum\limits_\alpha A_\alpha A_\alpha^*\leq I$, we have
$\sum\limits_\alpha |\lambda_{k,\alpha}|^2\leq 1$ for every $k$. Let
$\xi_k=\{\lambda_{k,\alpha}\}_{\alpha\in \Lambda}\in l^2(\Lambda)$,
then $\|\xi_k\|\leq 1$ for every $k$. Thus if
$\langle\xi_{k_1},\xi_{k_2}\rangle=1$, then by Schwarz inequility we
have  $\xi_{k_1}=\xi_{k_2}$. So by the assumption above, we conclude
that $k_1=k_2$.

Now we suppose $B\in B(H)^{\Phi_{\Gamma}}$. Then
$B=\sum\limits_\alpha A_\alpha B A_\alpha^*$. So
$P_kBP_l=(\sum\limits_\alpha
\lambda_{k,\alpha}\overline{\lambda_{l,\alpha}})P_kBP_l=\langle\xi_{k},\xi_{l}\rangle
P_kBP_l$ for every $k,l$. Thus we have $P_kBP_l=0$ for $k\neq l$. So
$B=\sum\limits_{k} P_kBP_k$. We conclude that $BP_k=P_kB$ and $B\in
{\Gamma}'$. That is, $B(H)^{\Phi_{\Gamma}}\subseteq {\Gamma}'$.

\vskip 0.2 in

{\bf 3. Almost sharp quantum effects}

\vskip 0.2 in

Firstly, let ${\cal E}(H)$ be the set of self-adjoint operators on
$H$ satisfying that $0\leq A\leq I$. For $A\in B(H)$, denote
$Ker(A)=\{x\in H\mid Ax=0\}$ and $Ran(A)=\{Ax\mid x\in H\}$. If
$A,B\in {\cal E}(H)$, we call $A\circ
B=A^{\frac{1}{2}}BA^{\frac{1}{2}}$ the sequential product of $A$ and
$B$ (see [7-10]).

\vskip 0.1 in

{\bf Lemma 3.1 ([7-8]).} If $A,B\in {\cal E}(H)$, $A\circ B\in
P(H)$, then $AB=BA$.

\vskip 0.1 in

We generalize Corollary 3 in [3] as the following Theorem 3.1.

{\bf Theorem 3.1.} Suppose $P\in P(H)$, $A\in {\cal E}(H)$, $P\ or\
A\in T(H)$, then the following conditions are all equivalent:

(1) $P\circ A\in P(H)$;

(2) $tr(PA)=tr(PAPA)$;

(3) $PA\in P(H)$;

(4) $PA$ is idempotent.

{\bf Proof.} (1)$\Rightarrow$(3). By Lemma 3.1 we have $PA=AP$. Thus
$PA=PAP=P\circ A\in P(H)$.

(3)$\Rightarrow$(4)$\Rightarrow$(2) is obvious.

(2)$\Rightarrow$(1). Since $P\circ A\in T(H)$, we have $(P\circ
A)^2\in T(H)$.

$tr(P\circ
A)=tr(PAP)=tr(PA)=tr(PAPA)=tr(PAPAP)=tr((PAP)^2)=tr((P\circ A)^2)$.

Since $0\leq P\circ A\leq I$, we have $(P\circ A)^2\leq P\circ A$.
It follows from $tr(P\circ A-(P\circ A)^2)=0$ that $P\circ A=(P\circ
A)^2$. So $P\circ A\in P(H)$.

\vskip 0.2 in

Let $M$ be a von Neumann algebra on $H$. The set of effects in $M$
is ${\cal E}(M)=\{A\in M\mid 0\leq A\leq I\}$. The set of
projections or sharp effects in $M$ is $P(M)=\{P\in M\mid
P=P^*=P^2\}$. We denote the usual Murray-von Neumann relations on
$P(M)$ by $\preceq$, $\succeq$ and $\sim$.

For $A\in {\cal E}(M)$, defining the {\it negation} of $A$ by
$A'=I-A$. if $A=PQP$ for some $P,Q\in P(M)$, we say
  $A$ is an {\it almost sharp} element in $M$. We
say that $A$ is {\it nearly sharp} if both $A$ and $A'$ are almost
sharp ([3]).

We denote the set of almost sharp elements in $M$ by $M_{as}$.

For $A\in {\cal E}(M)$, we denote the projection onto
$\overline{Ran(A)}$ and $Ker(A)$ by $P_A$ and $N_A$ respectively. It
is easy to know that $P_A+N_A=I$.

Note that if $A\in \varepsilon(M)$ has the form $A=PQP$ for some
$P,Q\in P(M)$, then $P_A\leq P$, thus we also have that $A=P_AQP_A$
([3]).

\vskip 0.1 in

{\bf Lemma 3.2 ([3]).} Let $A\in {\cal E}(M)$. Then

(1). $A$ is almost sharp iff $P_{AA'}\preceq N_A$;

(2). $A$ is nearly sharp iff $P_{AA'}\preceq N_A$ and
$P_{AA'}\preceq N_{A'}$;

(3). $P_{AA'}=P_A-N_{A'}=I-N_A-N_{A'}$.

\vskip 0.1 in

Now, we generalize Theorem 10 in [3] as the following Theorem 3.2
and Theorem 3.3:

{\bf Theorem 3.2.} Suppose $P\in P(M)$, then the following
conditions are all equivalent:

(1). $P\preceq P'$;

(2). $[0,P]\subseteq M_{as}$.

{\bf Proof.} (1)$\Rightarrow$(2). Suppose $0\leq A\leq P$. Then
$P_A\leq P$, $N_A\geq P'$. Thus $P_{AA'}\leq P_A\leq P\preceq P'\leq
N_A$. That is, $P_{AA'}\preceq N_A$. So by Lemma 3.2 we have $A\in
M_{as}$.

(2)$\Rightarrow$(1). Let $A=\frac{1}{2}P$, then $A\in [0,P]\subseteq
M_{as}$. So by Lemma 3.2 we have $P_{AA'}\preceq N_A$.

It is easy to see that $P_A= P$, $N_A= P'$, $N_{A'}= 0$. By Lemma
3.2 we have $P_{AA'}=P_A-N_{A'}=P$. Thus $P=P_{AA'}\preceq N_A=P'$.

\vskip 0.1 in

{\bf Theorem 3.3.} Suppose $P\in P(M)$, then the following
conditions are all equivalent:

(1). $P\sim P'$;

(2). $[0,P]\cup[0,P']\subseteq M_{as}$;

(3). If $A\in {\cal E}(M)$, $AP=PA$, then $A=P_1Q_1P_1+P_2Q_2P_2$
with $P_i, Q_i\in P(M)$ and $P_1\leq P$, $P_2\leq P'$.

{\bf Proof.} (1)$\Longleftrightarrow$(2). By Theorem 3.2.

(2)$\Rightarrow$(3). Suppose $A\in {\cal E}(M)$, $AP=PA$. Then
$A=PAP+P'AP'$. Since $PAP\in [0,P]$ and $P'AP'\in [0,P']$, we have
$PAP,P'AP'\in M_{as}$. Thus, we can prove the result easily.

(3)$\Rightarrow$(2). Suppose $0\leq A\leq P$. It is easy to see that
$AP=PA=A$. Thus $A=P_1Q_1P_1+P_2Q_2P_2$ with $P_i, Q_i\in P(M)$ and
$P_1\leq P$, $P_2\leq P'$. So $A=PAP=P_1Q_1P_1$. That is, $A\in
M_{as}$. We conclude that $[0,P]\subseteq M_{as}$. Similarly
$[0,P']\subseteq M_{as}$.

\vskip 0.2 in

Let ${\cal B}[0,1]$ be the set of bounded Borel functions on
interval $[0,1]$. Suppose $A\in {\cal E}(M)$, $h\in {\cal B}[0,1]$,
$0\leq h\leq 1$, then $h(A)\in
 {\cal E}(M)$.

\vskip 0.2 in

{\bf Theorem 3.4.} Suppose $A\in {\cal E}(M)$, $h\in {\cal B}[0,1]$,
$0\leq h\leq 1$, $h(0)=0$, $h(1)=1$. We have

(1). $N_A\leq N_{h(A)}$, $N_{A'}\leq N_{h(A)'}$, $P_{h(A)h(A)'}\leq
P_{AA'}$;

(2). If $A$ is almost sharp, then $h(A)$ is almost sharp;

(3). If $A$ is nearly sharp, then $h(A)$ is nearly sharp.

{\bf Proof.} (1). If $Ax=0$, then $h(A)(x)=h(0)x=0$. Thus
$Ker(A)\subseteq Ker(h(A))$. That is, $N_A\leq N_{h(A)}$.

If $Ax=x$, then $h(A)(x)=h(1)x=x$. Thus $Ker(I-A)\subset
Ker(I-h(A))$.  That is, $N_{A'}\leq N_{h(A)'}$. Thus by Lemma 3.2 we
have $P_{AA'}=I-N_A-N_{A'}\geq I-N_{h(A)}-N_{h(A)'}=P_{h(A)h(A)'}$.

(2). If $A$ is almost sharp, by Lemma 3.2 we have $P_{AA'}\preceq
N_A$. From (1) we have $P_{h(A)h(A)'}\leq P_{AA'}\preceq N_A\leq
N_{h(A)}$.  That is, $P_{h(A)h(A)'}\preceq N_{h(A)}$. Thus by Lemma
3.2 again $h(A)$ is almost sharp.

(3). If $A$ is nearly sharp, by Lemma 3.2 we have $P_{AA'}\preceq
N_A$ and $P_{AA'}\preceq N_{A'}$. From (1) we have
$P_{h(A)h(A)'}\leq P_{AA'}\preceq N_A\leq N_{h(A)}$ and
$P_{h(A)h(A)'}\leq P_{AA'}\preceq N_{A'}\leq N_{h(A)'}$.  That is,
$P_{h(A)h(A)'}\preceq N_{h(A)}$ and $P_{h(A)h(A)'}\preceq
N_{h(A)'}$. Thus by Lemma 3.2 again $h(A)$ is nearly sharp.

\vskip 0.2 in

Let $C[0,1]$ be the set of continuous functions on interval $[0,1]$.
Suppose $h\in C[0,1]$, we say $h$ satisfy {\it kernel condition} if
the following three conditions hold:

(1). $0\leq h\leq 1$;

(2). $h(0)=0$, $h(1)=1$;

(3). $h$ is strictly monotonous.

\vskip 0.2 in

Suppose $A\in {\cal E}(M)$, $h\in C[0,1]$ satisfies kernel
condition, then it is easy to see that $h(A)\in {\cal E}(M)$,
$h^{-1}\in C[0,1]$ also satisfies kernel condition and
$A=h^{-1}(h(A))$.

\vskip 0.2 in

{\bf Theorem 3.5.} Suppose $A\in {\cal E}(M)$, $h\in C[0,1]$ satisfy
kernel condition. We have

(1). $N_A=N_{h(A)}$, $N_{A'}=N_{h(A)'}$, $P_{AA'}=P_{h(A)h(A)'}$;

(2). $A$ is almost sharp if and only if $h(A)$ is almost sharp;

(3). $A$ is nearly sharp if and only if $h(A)$ is nearly sharp.

{\bf Proof.} (1). By Theorem 3.4, we have $N_A\leq N_{h(A)}$,
$N_{A'}\leq N_{h(A)'}$, $P_{h(A)h(A)'}\leq P_{AA'}$. Since $h(A)\in
 \varepsilon(M)$, $h^{-1}\in C[0,1]$ satisfy kernel
condition, and $A=h^{-1}(h(A))$, by Theorem 3.4 again, we have
$N_A\geq N_{h(A)}$, $N_{A'}\geq N_{h(A)'}$, $P_{h(A)h(A)'}\geq
P_{AA'}$. Thus the conclusion follows.

(2) and (3) follow from Lemma 3.2 and (1) immediately.

\vskip 0.1 in

{\bf Corollary 3.1.} Suppose $A\in {\cal E}(M)$, $t$ is a positive
number. Then

(1). $A$ is almost sharp if and only if $A^t$ is almost sharp.

(2). $A$ is nearly sharp if and only if $A^t$ is nearly sharp.

\vskip0.2in

\centerline{\bf References}

\vskip0.2in

\noindent [1]. Nielsen, M. and Chuang, J. {Quantum computation and
quantum information}, Cambridge University Press, 2000

\noindent [2]. A. Arias, A. Gheondea, S. Gudder. Fixed points of
quantum operations. J. Math. Phys. 43(2002), 5872.

\noindent [3]. A. Arias, S. Gudder. Almost sharp quantum effects. J.
Math. Phys. 45(2004), 4196.

\noindent [4]. M. D. Choi. A Schwarz inequality for positive linear
maps on $C^*$-algebras. Illinois J. Math. 18(1974), 565.

\noindent [5]. R. V. Kadison.  A generalized Schwarz inequality and
algebraic invariants for operator algebras. Ann. of Math. 56(1952),
494.

\noindent [6]. R. Horn, C. Johnson. Matrix Analysis. Cambridge
University Press, Cambridge, 1990.

\noindent [7]. S. Gudder, G. Nagy. Sequential quantum measurements.
J. Math. Phys. 42(2001), 5212.

\noindent [8]. S. Gudder, R. Greechie.  Sequential products on
effect algebras. Rep. Math. Phys.  49(2002), 87.

\noindent [9]. Weihua Liu, Junde Wu. A uniqueness problem of the
sequence product on operator effect algebra ${\cal E}(H)$. J. Phys.
A: Math. Theor. 42 (2009), 185206.

\noindent [10]. Jun Shen, Junde Wu. Sequential product on standard
effect algebra ${\cal E}(H)$. J. Phys. A: Math. Theor. 44 (2009).

\vskip0.2in

\centerline{\bf Appendix: Collected papers of Shen Jun}

\vskip0.2in

\noindent [1]. Jun Shen, Junde Wu. Not each sequential effect
algebra is sharply dominating. Physics Letters A. 373 (2009),
1708-1712.

\noindent [2]. Jun Shen, Junde Wu. Sequential product on standard
effect algebra ${\cal E}(H)$. J. Phys. A: Math. Theor. 44 (2009).

\noindent [3]. Jun Shen, Junde Wu. Remarks on the sequential effect
algebras. Reports on Math. Phys. 63 (2009), 441-446.

\noindent [4]. Jun Shen, Junde Wu. The average value inequality in
sequential effect algebras. Acta Math. Sinica, English Series.
Accepted for publishing.

\noindent [5]. Jun Shen, Junde Wu. The n-th root of sequential
effect algebras. Submitted.

\noindent [6]. Jun Shen, Junde Wu. Spectral representation of
infimum of bounded quantum observables. Submitted.

\noindent [7]. Jun Shen, Junde Wu. Generalized quantum operations
and almost sharp quantum effects. Submitted.

\end{document}